# One neuron is more informative than a deep neural network for aftershock pattern forecasting

Arnaud Mignan[1,2], Marco Broccardo[2]

[1] *Swiss Federal Institute of Technology, Institute of Geophysics, Zurich*
[2] *Swiss Seismological Service, Zurich*

*Version date: 3 April 2019*

29 August 2018: "*Artificial intelligence nails predictions of earthquake aftershocks*[1]". This Nature News headline is based on the results of DeVries et al.[2] who forecasted the spatial distribution of aftershocks using Deep Learning (DL) and static stress feature engineering. Using receiver operating characteristic (ROC) curves and the area under the curve (AUC) metric, the authors found that a deep neural network (DNN) yields AUC = 0.85 compared to AUC = 0.58 for classical Coulomb stress. They further showed that this result was physically interpretable, with various stress metrics (e.g. sum of absolute stress components, maximum shear stress, von Mises yield criterion) explaining most of the DNN result. We here clarify that AUC ≈ 0.85 had already been obtained using ROC curves for the same scalar metrics and by the same authors in 2017[3]. This suggests that DL—in fact—does not improve prediction compared to simpler baseline models. We reformulate the 2017 results[3] in probabilistic terms using logistic regression (i.e., one neural network node) and obtain AUC = 0.85 using 2 free parameters versus the 13,451 parameters used by DeVries et al.[2] We further show that measured distance and mainshock average slip can be used instead of stress, yielding an improved AUC = 0.86, again with a simple logistic regression. This demonstrates that the proposed DNN so far does not provide any new insight (predictive or inferential) in this domain.

Operational aftershock forecasting has been possible for decades thanks to well-established empirical laws [4-6]. Spatial patterns of aftershocks are often described as a power-law decay[6-9]. Physical models based on the Coulomb stress paradigm have been shown to perform worse than statistical methods on their own[10] but to outperform statistical methods when considered in physical/statistical hybrids and only when high-quality mainshock rupture data is available[11]. Meade et al.[3] performed a thorough analysis of various scalar stress metrics and showed that several outperform classic Coulomb failure stress. We are here concerned with their follow-up article[2] that presented similar results but via DL[1]. In this comment, we aim to clarify two important aspects of DL in the context of aftershock pattern prediction: (1) While defining larger and deeper DNNs usually does not hurt model performance, it decreases model interpretability for physical inference—we will show that a single neuron provides similar results as DeVries et al.'s DNN; (2) DL learns patterns from data, and it is preferable to use measurable observations as input features instead of model-derived data (e.g. stress) for disambiguation.

We first reproduce the results of DeVries et al.[2], as illustrated in Fig. 1a-c (see their method section). Their model contains 13,451 free parameters (weights and biases) since their DNN is made of 6 layers each composed of 50 nodes. Their inputs are 12 engineered stress components: the absolute values of the tensor's 6 independent components $|\sigma_{ij}|$ and their 6 opposites $-|\sigma_{ij}|$. Their output is the probability that a spatial cell is in binary class $y$ (1 if aftershocks are present, 0 otherwise). We retrieve an AUC = 0.85 and a precision of 5.4% at a threshold $p$ = 0.5 as originally published (notice that the slightly different precision is due to the random subsampling of the balanced training dataset). However, a similar AUC can be obtained when using only one input (scalar stress metric) and one node (2 free parameters with one weight $\beta_1$ and one bias $\beta_0$). This is illustrated in Fig. 1d-f where we introduce a logistic regression for direct comparison with DeVries et al.'s DNN. The classifier is defined as $\Pr(y) = 1/(1+\exp[-(\beta_0+\beta_1 \log_{10}(x)]))$ where $y$ is the spatial cell class and $x$ the chosen scalar stress metric. We obtain AUC = 0.85 and a precision of 5.4% at a threshold $p$ = 0.5 for the sum of absolute stress components (Fig. 1d-f), maximum shear stress, and von Mises yield criterion.

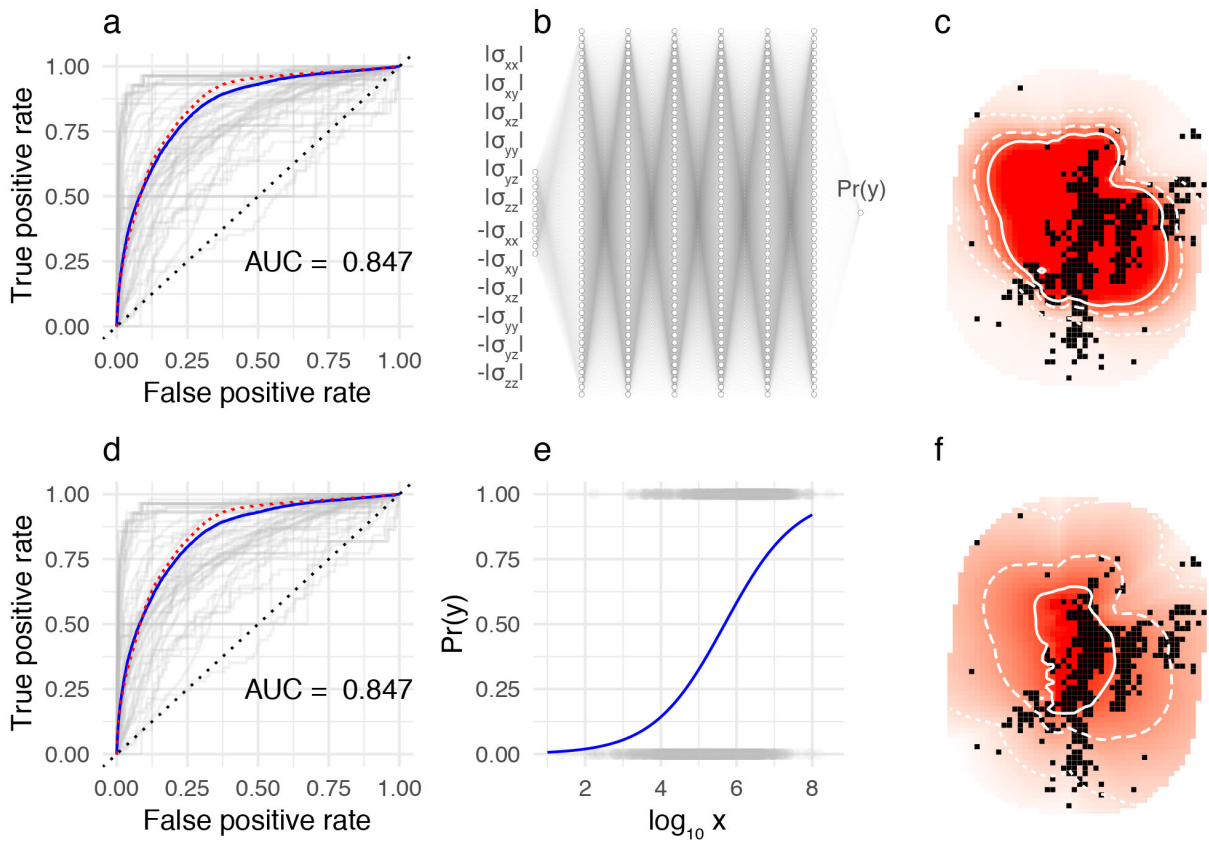

**Fig. 1 | Prediction of aftershock spatial patterns based on stress features. a.** Test data ROC curves for DeVries et al.'s DNN; **b.** DNN topology (plot generated with alexlenail.me/NN-SVG); **c.** Example of DNN prediction (1999 ChiChi aftershocks); **d.** Test data ROC curves for logistic regression with log of the sum of absolute stress components as input; **e.** Logistic regression fit on training data; **f.** Example of logistic regression prediction. For both models, 58 ROC curves are shown, for the 57 mainshocks from the test set in grey and all combined mainshock-aftershock pairs in blue (see Fig. 2 for red color explanation). Dotted, dashed and solid curves in c,f represent $\Pr(y)$ = 0.3, 0.5 and 0.7, respectively.

DL should be used directly on observable and measurable variables, avoiding or reducing *de facto* feature engineering and hidden assumptions. However, DeVries and colleagues used a combination of the stress tensor components as DNN input. The stress tensor is not measured but estimated on the basis of different assumptions (e.g. homogeneous medium, linearized elasticity theory), and measured or assumed quantities (e.g. distributions of rupture slip, Lamé constants, friction coefficient, regional stress), some of which are affected by large uncertainties[12]. If not properly quantified, these uncertainties significantly influence local stress calculations, limiting the overall quality of any stress-based binary classifier (even for a complex DNN). In fact, accuracy seems to reach an AUC plateau of 0.85-0.86. A different and simpler approach—for the same binary classification approach—is to use measurable variables, which are less affected by these uncertainties.

In the following, we show that a logistic regression based on mainshock average slip, $d$, and minimum distance $r$ between space cells and mainshock rupture (i.e. the simplest of the possible models, with orthogonal features), provides comparable or better accuracy than a DNN. Both $d$ and $r$ [m] were obtained from the SRCMOD database[2]. Results are shown in Fig. 2. The performance is improved to AUC = 0.86 and same precision 5.4% compared to the ones obtained with stress features[2-3]. The distance-slip probabilistic model is described by

$$\Pr(y) = 1/\left(1 + e^{-[\beta_0 + \beta_1 \log_{10}(r) + \beta_2 \log_{10}(d)]}\right) \qquad (1)$$

where $\beta_0 = 10.18 \pm 0.07$, $\beta_1 = -2.32 \pm 0.02$ and $\beta_2 = 1.16 \pm 0.01$. Eq. (1) provides a transparent and interpretable model to forecast aftershock patterns from geometric and kinematic data, retrievable in near real-time after a mainshock. This approach is in line with both the literature on operational earthquake forecasting[4-6] and statistical seismology[7-9].

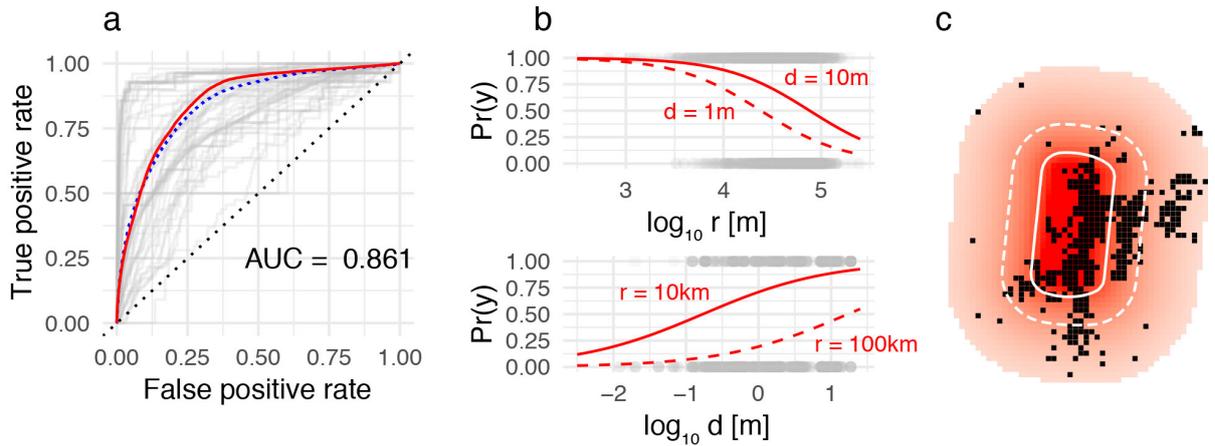

**Fig. 2 | Prediction of aftershock spatial patterns based on distance $r$ and slip $d$. a.** Test data ROC curves for the logistic regression; **b.** Logistic regression fit on training data; **c.** Example of logistic regression prediction. 58 ROC curves are shown, for the 57 mainshocks of the test set in grey and all combined mainshock-aftershock pairs in red (see Fig. 1 for blue color explanation). Dashed and solid curves in c represent $\Pr(y) = 0.5$ and 0.7, respectively.

This communication shows that—given the same datasets and same accuracy assessments proposed by DeVries et al.[2]—DL does not offer new insights or better accuracy in predicting aftershock patterns. However, we strongly believe that DL is revolutionizing data analytics in many domains[13-14], including statistical seismology[15]. Therefore, the objective of our study is not to restrain the use of DL in this field, but to stimulate a further research effort.